\documentclass[twocolumn,aps,prl,superscriptaddress]{revtex4}%
\usepackage{color}
\usepackage{amsmath}
\usepackage{amsfonts}
\usepackage{amssymb}
\usepackage{hyperref}
\usepackage{graphicx}%
\providecommand{\U}[1]{\protect\rule{.1in}{.1in}}
\newcommand{\ket}[1]{{|#1\rangle}}

\begin{document}
\title{Weak randomness {completely trounces} the security of QKD}

\author{{Jan Bouda}} \affiliation{Faculty of
Informatics, Masaryk University, Brno, Czech
Republic}\author{{Matej Pivoluska}} \affiliation{Faculty of
Informatics, Masaryk University, Brno, Czech Republic}
\author{{Martin Plesch}}
\affiliation{Faculty of Informatics, Masaryk University, Brno,
Czech Republic} \affiliation{Institute of Physics, Slovak Academy
of Sciences, Bratislava, Slovakia}
\author{{Colin Wilmott}}
\affiliation{Faculty of Informatics, Masaryk University, Brno,
Czech Republic}

\begin{abstract}
In usual  {security proofs} of quantum protocols the adversary
(Eve) is expected to have full control {over} any quantum
communication between any communicating parties (Alice and Bob).
Eve is also expected to have full access to {an} authenticated
classical channel between Alice and Bob. Unconditional security
against any attack {by} Eve can be proved even in the realistic
setting of device and channel imperfection. In this Letter we show
that the security of QKD protocols is ruined if one allows Eve to
{possess a very limited} access to {the random sources} used by
Alice. Such {knowledge} should always  be expect{ed} in realistic
experimental conditions via different side channels.

\end{abstract}
\maketitle

\emph{Introduction ---} The emergence of quantum theory in the
early twentieth century led to a revolution in many areas of
physics. One of its main features was the introduction of
intrinsic randomness, originating from the very nature of the
theory. This probabilistic nature led to questioning of concepts
of (macro)realism and locality
\cite{EinsteinPodolskyRosen-CanQuantum-MechanicalDescription-1935}
{which} was considered as an unwanted consequence of quantum
theory. True randomness, much   unwanted from the point of view of
classical physics, serves as a valuable resource in many
cryptographic protocols. {It is for this reason that} {q}uantum
random number generators (QRNG) were one of the first commercially
available devices utilizing basic principles of quantum physics in
its elementary nature.


Towards the latter part of the twentieth century it was recognized
that quantum mechanics could lead another revolution and
dramatically extend the premise of information processing.
Classical notions of security underpinned by computational
conditions were seriously threatened by the edicts of quantum
mechanics and by {the emergence of} Shor's algorithm
\cite{Shor-Polynomial-TimeAlgorithmsPrime-1997}. However, quantum
mechanics offered a new security paradigm whereby the use of
quantum states imparted unconditional secure communication through
\emph{quantum key distribution (QKD)} \cite{BB84}. {Q}uantum key
distribution protocol{s} {enable} two communicating parties to
produce a shared random secret key in such a way that also reveals
the presence of any third party. The secret key can  be used later
to implement an unconditionally secure encryption protocol
\cite{Vernam-CipherPrintingTelegraph-1926}.

The security of QKD has been established not only for an ideal
noiseless experimental setting, but it has {also been} prove{n}
robust within more realistic settings to the exten{t} that QKD
systems are now commercially available \cite{QNRG}. Interestingly,
the robustness of  QKD protocols has only been proven with respect
to possible attacks on quantum data exchanged by the communicating
parties with the assumption that a third party possesses knowledge
of all exchanged classical data.

Sources of classical random bits, repeatedly used during different
phases of quantum protocols, were silently considered being
perfect. An unstated assumption in the standard proofs of security
\cite{Lo1999,Mayers2001,Shor2000} is that the source of random
bits used in the protocol is unbiased and completely unaccessible
 to the adversary. Unfortunately, however, perfect or unbiased
randomness is very difficult to obtain in practice. All classical
sources of random bits provide in fact only pseudorandom bit
strings,  which might be fully accessible to the adversary
together with knowledge of its preparation procedure and input
bits. Specialized QRNG devices only produce   biased randomness
and require classical post processing
\cite{Solcgravea-TestingofQuantum-2010}, something one has to
consider as accessible to the adversary. {Real} world random
number generators inevitably leak information via side channels
{and, thus, may} be vulnerable to outside conditions (e.~g.,
temperature, input power, EM radiation etc.)~{which are}
potentially controlled by the adversary.

Although the problem of weak (biased) randomness has been broadly
studied and is relatively well understood in classical information
processing
\cite{DodisPrabhakaran-(im)possibilityofcryptography-2004,MaurerWolf-PrivacyAmplificationSecure-1997,
McInnes,RennerWolf-UnconditionalAuthenticityand-2003}, there has
not been a similar analysis of its quantum counterpart. This {may}
be due to the fact that there theoretically exist{s} a perfect
source of randomness in quantum world {and any} weakness{es} {are}
only {attributed} to imperfect implementation. Recent
investigations however {show} that quantum information processing
can help to increase security of communication using weak
randomness even for regions of parameters where  purely classical
processing would inevitably reveal all information to the
adversary \cite{enc2qbit,
BoudaPivoluskaPlesch-EncryptiontoMultiple-2012}.


In this Letter we will examine the {security setting} of {QKD in}
which the adversary, aside from having a full control of the
quantum and classical channel, has also some limited control over
the sources of randomness the communicating parties employ during
the protocol (Fig. 1). We will show that with {an} increasing key
length, only a negligible control of the randomness is necessary
to render the QKD insecure. In particular, we will demonstrate
that the secret key individually held by communicating parties
will {differ significantly.} Moreover, { knowledge pertaining to
the secret key held by the adversary will be
 comparable to the knowledge held by the receiving party.}

\begin{figure}[ptb]
\begin{center}
\includegraphics[scale=0.37]
{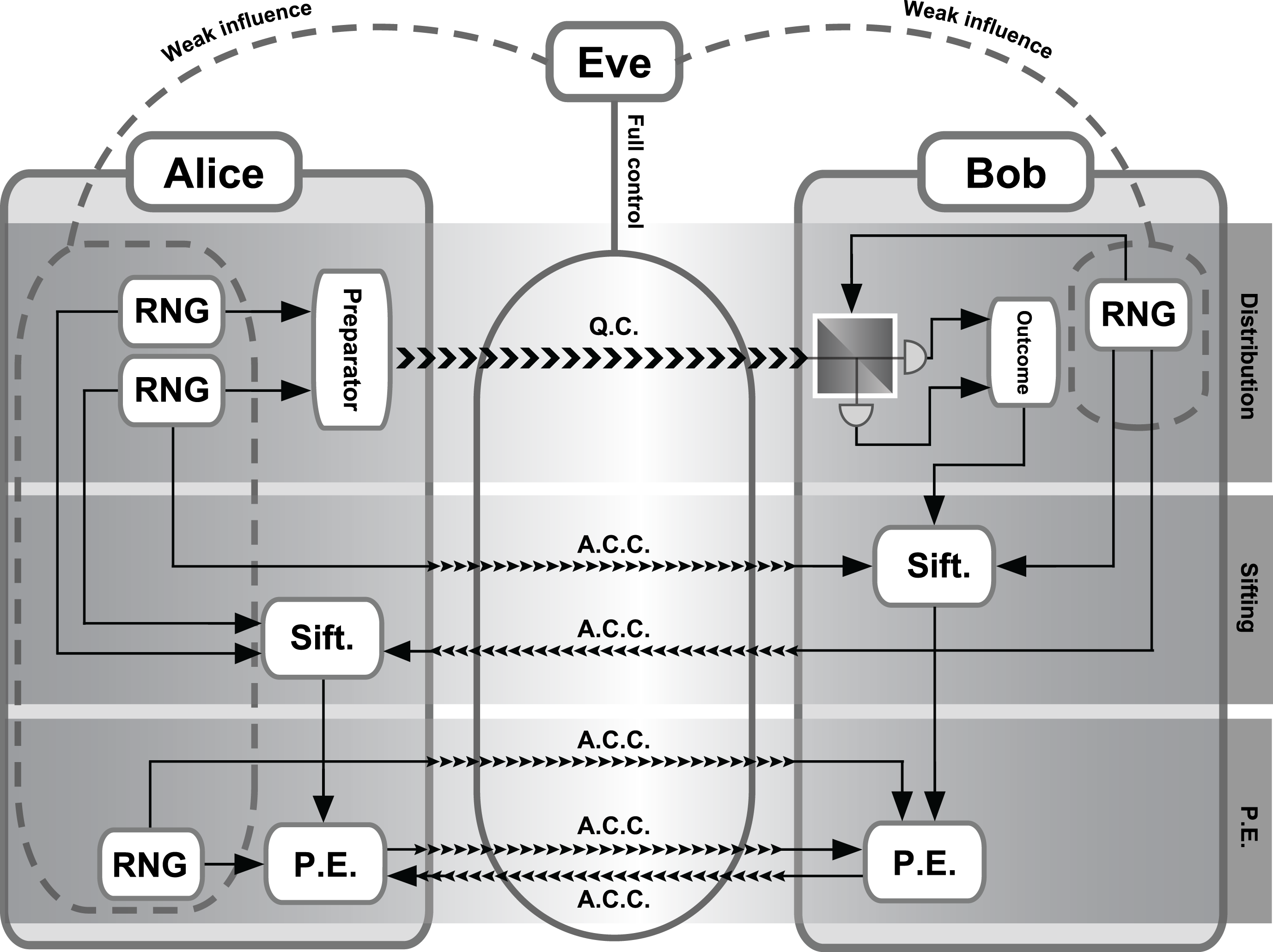}
\end{center}
\caption{A sketch   version of the BB84 protocol. Eve has full
access to  both quantum channel (Q.C.) and to authenticated
classical channel (A.C.C.), and possesses a partial
access to random sources of Alice and Bob.}%
\label{figure1}%
\end{figure}


\emph{Weak sources ---} Random processes are usually described by
their probability distributions. However, it is insufficient to
model a weak random source by a single probability {distribution
because} the bias of the source is typically unknown. The only
information usually known about the source is that it is {random
to} a certain extent; thus, we allow the output of the weak random
source to be distributed according to any probability distribution
containing {sufficient} randomness. We {will} quantify the amount
of randomness of a distribution by {the} $\emph{min--entropy}$ of
its source. The min--entropy of the random variable $\mathbf{X}$
is defined by
\begin{equation}
H_{\infty}(\mathbf{X})=\min_{x\in X}\left(  -\log_2 Pr\left(  \mathbf{X}%
=x\right)  \right)  .\label{minentropy}%
\end{equation}
A  non--uniform source of randomness is an \emph{$(N,b)$-source}
if it emits $N$-bit strings drawn according to a probability
distribution with {a} min-entropy {of} at least $b$ bits. Thus,
every specific $N$-bit sequence is drawn with probability smaller
or equal to $2^{-b}$. For $b=N$, one {obtains} a perfect source
where all sequences are drawn with the same probability. The bias
of the source can be easily quantified by the \emph{min-entropy
loss} denoted $c=N-b$. A distribution is \emph{$(N,b)$--flat} if
it is an $(N,b)$--source and it is uniform on a subset of $2^{b}$
sample points, i.~e., each string is {outputted} with {a}
probability of either zero or $2^{-b}$.


{The} {q}uantity $\frac{b}{N}$ is called {the} \emph{min--entropy
rate} {and it achieves unity}   for perfect random sources {that
deliver} one bit of entropy per bit produced. We will be
particularly interested in the {\emph{min--entropy loss rate}
which will be denoted by quantity $\frac{c}{N}$}.  This quantity
is (almost) zero for (almost) perfect random sources and
{approaches unity} {as the quality of the source  decreases}.


\emph{The QKD protocol ---} Here we demonstrate the attack using a
variation of the  {well-known} BB84 protocol \cite{BB84} {which
serves} as a representative {for} the prepare-measure family of
protocols.

\textit{Distribution phase:}  {Using a random number generator
Alice produces a $2n$--bit string $X$. Then depending on a
$2n$--bit string from a random variable $Y$, Alice encodes each
bit into a qubit from one of four possible states
$\{\ket{0},\ket{1}$, $\ket{+},\ket{-}\}$. The state of the $i^{\rm
th}$-qubit is conditioned on the $i^{\rm th}$-bits of both $X$ and
$Y$. In particular, each bit of $X$ with value 0 is encoded into
either $\ket{0}$ or $\ket{+}$ depending on whether the
corresponding bit of $Y$ is 0 or 1 respectively. A similar case
holds for the bit $1$ encoding into the states $\ket{1}$ and
$\ket{-}$. Alice subsequently transmits all $2n$--qubits to Bob.
In order to obtain information about the $2n$--bit string $X$, an
adversary will be compelled to interact with these transmitted
qubits which inevitably leads to a disturbance in the transmitted
sequences.}

\textit{Sifting phase:} {Bob measures each received qubit to
obtain a $2n$--bit string. {Similar} to the encoding procedure, a
set of measurement bases are chosen according to a uniformly
distributed random variable $Z$ that outputs a $2n$--bit string.
If the $i^{\rm th}$--bit of $Z$ has value 0, Bob measures in the
computational basis otherwise Bob measures in the diagonal basis.
The sequence of measurement bases is revealed by Bob whereupon
Alice then announces the locations of those qubits for which the
corresponding preparation  and measurement bases do no coincide.
After discarding these qubits, Alice and Bob  possess on average
  $n$--bit strings $X_{A}$ and $X_{B}$. Following the sifting
phase, the adversary has an estimate $X_{E}$ of Alice's string
$X_{A}$ that depends on the degree to which the adversary
interacted with the transmitted qubits. If there is no interaction
then the adversary possesses no information on the $n$--bit string
$X_{A}$. In the case of faultless quantum communication, $X_A$ and
$X_B$ will be identical. However, in the case of the adversary
choosing to interact with many qubits, the estimate $X_E$ will be
a good approximation to $X_A$, and this causes $X_B$ to differ
significantly from $X_A$.}

\textit{Parameter estimation:} {The primary aim of parameter
estimation is to approximate the number of errors between the
$n$--bit strings $X_A$ and $X_B$. The source of the  errors may be
attributed to a combination of quantum channel imperfections or
eavesdropping by the adversary. However, in security proofs, one
always considers the worst case scenario and, thus, assumes the
adversary to be responsible for all errors.}

 {Random sampling provides a way to estimate the number of errors
between $X_A$ and $X_B$. According to the output from a random
variable $T$, Alice chooses a set of bit positions of $X_{A}$ and
assigns these as the test positions. Alice and Bob   reveal the
bit value in each test position. The  number of   errors $t$
provides a reasonable estimate $r$ on the actual number of errors
in the remaining bits of $X_{A}^{R}$ and $X_{B}^{R}$
\cite{Mayers2001}. If the number of errors in the test positions
is excessive then there is a high probability that the adversary
is present and the protocol is aborted.}

In any practical application one wants the test set to be
relatively small in order to achieve a maximal  possible  key
length. In existing QKD protocols,  the size of the test set is
typically in order of $\sqrt{n}$ or $\log(n)$
\cite{christandl-2004,HorodeckiHorodeckiHorodeckiEtAl-QuantumKeyDistribution-2008,
LoChauArdehali-EfficientQuantumKey-2005}. In the following
asymptotic analysis, we assume the most general case and post only
a condition that the size of the test set is sublinear in $n$. In
particular{, we assume} it is equal to $\Theta (n^{1-\alpha})$
with $0<\alpha<1$.

\textit{Information reconciliation and privacy amplification:}
{Following  parameter estimation, the bit strings $X_{A}^{R}$ and
$X_{B}^{R}$ contain with a high probability up to $r$ errors. The
goal of the information reconciliation is to remove these errors
even at the cost of revealing some information about $X_{A}^{R}$
and $X_{B}^{R}$.}   This task is usually realized by one--way
communication
\cite{LoChauArdehali-EfficientQuantumKey-2005,Mayers2001,Shor2000}.
Such one way information reconciliation can be implemented as long
as Bob has more information {than the adversary} about Alice's
string $X_{A}^{R}$
\cite{CsiszarKorner-Broadcastchannelswith-1978}.

The goal of the privacy amplification is to remove any knowledge
{possessed by Eve} about the shared string $X_{A}^{R}$. A widely
used method
\cite{BennettBrassardCrepeauEtAl-Generalizedprivacyamplification-1995,Renner2005}
is based on the random choice of a hashing function. {In this
case,} Alice randomly chooses a hashing function $f$ and sends it
to Bob. The final shared key is $f(X_{A}^{R}) = f(X_{B}^{R})$.
Importantly, this method {also  uses} one--way communication.

\emph{The Adversary's attack ---} {The use of uniform randomness
is widespread throughout the various steps of the QKD protocol.
The first instance of uniform randomness occurs during the
distribution phase when Alice chooses $2n$--bit strings $X$ and
$Y$ uniformly. During the sifting phase of the protocol, Bob must
decide on a set of measurement bases which is again dependent upon
a uniformly distributed random variable that outputs a $2n$--bit
string. In the parameter estimation phase, a subset of the strings
is chosen as a test set according to a uniformly distributed
random variable $T$ and, again, another source of random bits is
used to select the hashing function. In light of these cases, we
will investigate a scenario in which Alice's randomness source  -
used to select the positions of test qubits -} is biased. Similar
to the case of faulty quantum channels we will consider the worst
case scenario and {attribute} all randomness imperfections to the
adversary. This can be modelled by a {scenario whereby}   the
adversary can influence the distribution of the random variable
$T$ to such an extent that the adversary is allowed to set any
$(n,n-c)$--distribution to the random variable $T$ {with $c$
denoting the strength} of the adversary's attack. We   assume that
$c$ is large enough {to guarantee the adversary} that at least
half of the qubits will {not be tested}. Later we   calculate {the
required value of $c$.}

Without the loss of generality, let us suppose that the first half
of Bob's measurement outcomes will not be tested. The adversary
can measure the first half of the $2n$-qubits  in the $\{
|0\rangle , |1\rangle \}$ basis. If  Eve's measurement outcome is
${|0\rangle}$, she sends a state ${|1\rangle}$ to Bob and if her
 measurement  outcome {is} ${|1\rangle}$, she sends a state ${|0\rangle}$.
{Following this} procedure and the sifting phase, the adversary
has on average $\frac{n}{2}$ measurement outcomes. The adversary
adds another $\frac{n}{2}$ bits chosen randomly and uniformly to
obtain her estimate $X_{E}$ of Alice's string $X_{A}$. {Since
Alice and Bob have not tested those bits measured by the
adversary}, the protocol will continue on to remaining phases.

{We now quantify the amount of information that Bob and the
adversary possess about Alice's $n$--bit string $X_{A}$. To obtain
the result, we   calculate the Hamming distance $H(A,B)$ between
strings $A$ and $B$. There are three cases to consider. Firstly,
the adversary may have measured a transmitted qubit in the correct
basis. In such a case, the adversary obtains a bit value that
coincides with the corresponding bit value in $X_A$ with Bob then
obtaining the bit complement. This happens on average in ${n}/{4}$
measurement cases. Secondly, it may happen that the adversary
measures a transmitted qubit in incorrect basis. Here both Bob and
the adversary obtain the correct value with probability ${1}/{2}$.
This happens on average in ${n}/{4}$ bits. The final situation to
consider is the case in which the adversary does not perform a
particular qubit measurement. The adversary then chooses random
values for these bit positions and   correctly guesses the value
with probability ${1}/{2}$. In this situation, Bob's measurement
value is due to measuring in the correct basis and, thus, he
determines the value of Alice's bit with certainty. This last
situation occurs in ${n}/{2}$ of the bits.}

{The amount of information that Bob and the adversary possess
about Alice's string $X_{A}$ is given by $H(X_{B},X_{A})$ and
$H(X_{E},X_{A})$ respectively. Both of these quantities are on
average equal to ${5n}/{8}$. Consequently,  the adversary and Bob
possess on average the same level of knowledge about Alice's
string. As the subsequent steps of the protocol demand that only
Alice communicates information, it follows that with the
conclusion of the protocol,  the adversary and Bob continue to
share the same level of information about Alice's bit sting. This
illustrates that ultimately there can be no privacy between Alice
and Bob.}

\emph{The strength of the Adversary ---} {It remains to quantify
how much information in terms of min--entropy loss  the adversary
requires in order to prevent parameter estimation on half of the
bit positions.} Alice needs $\log{\binom{n}{n^{1-\alpha}}}$ bits
to specify $n^{1-\alpha}$ positions out of $n$. On the other hand,
the adversary wants Alice to choose the $n^{1-\alpha}$ test bits
only from $\frac{n}{2}$ of the positions. {Apparently,} the best
option for the adversary {- in terms of the smallest entropy loss
-} is to set
any  $\left(  \log{\binom{n}{n^{1-\alpha}}%
},\log{\binom{n/2}{n^{1-\alpha}}}\right)$-flat distribution to
Alice's random number generator. Such a distribution would
uniformly select test bits only within the pre-selected half of
all positions.

{Of particular importance here will be an analysis of the relative
behavior of two quantities; the first quantity is the length of
the test bit string
\[
N=\log{\binom{n}{n^{1-\alpha}}},%
\] and the second quantity is the
min--entropy loss
\[
c=\log{\binom{n}{n^{1-\alpha}}}-\log{\binom{n/2}{n^{1-\alpha}}}.%
\]}
{Both of these quantities diverge  since Alice demands an
increased level of randomness to choose the test bits from an ever
increasing set size.} Now,  the min--entropy loss rate $c/N$
expresses the amount of total randomness required  to restrict all
possible test bit positions within a prescribed subset of the
total bit set. We will show that the rate $c/N$, which is given as
\begin{equation}
\frac{c}{N}=\frac{\log{\binom{n}{n^{1-\alpha}}}-
\log{\binom{n/2}{n^{1-\alpha}%
}}}{\log{\binom{n}{n^{1-\alpha}}}},
\end{equation}
remains finite.

We will consider this expression in the limit of large $n$ as all
current security proofs for various QKD protocols {have} only been
proven in the asymptotic regime of infinite key length. In
evaluating the min--entropy loss rate $c/N$ in the limit of large
$n$, we will make use of the Stirling approximation of the
factorial function $\log \left( n!\right)=\left( n+1/2\right)
\log\left( n\right) -n$. Furthermore, we can approximate  the
quantity $c$ as $n^{1-\alpha}\log\left(2\right)+ O\left(
\log\left(n\right)\right)$ while the quantity $N$ can the
approximated to $n^{1-\alpha}\log\left(n\right) + O
\left(n^{1-\alpha}\right)$. The min--entropy loss
 rate $c/N$ in the limit of large $n$ can be evaluated to
\begin{equation}
\frac{c}{N} \approx \frac{1}{\log\left(n\right)}.\label{c/N}
\end{equation}

Under the assumption of  perfect randomness, all QKD protocols
have been proven to be perfectly secure in the limit of an
infinitely large key size. However, implementing perfect
randomness is  difficult. By relaxing the assumption of perfect
randomness to reflect real life conditions, Eq.~(\ref{c/N})
illustrates that QKD  no longer  remains robust. In particular, a
negligible control on the source of randomness renders QKD
insecure.


\emph{Entanglement based protocols ---} In these protocols
{\cite{Lo1999,Ekert1991}}, parties share entangled pairs of
photons and employ monogamy of entanglement to build up security.
A portion of these states {are} used to check the monogamy {- and,
thus, exclude the  presence of an adversary - while the remaining
states are used to perform the protocol itself.} The test pairs
are selected by a random source exactly in the same way as in the
prepare-measure based protocols. Having access to the random
source of the selecting {party,} Eve might easily perform an
attack where she {c}ould entangle herself to pairs not being
tested in the future and{, thereby,} obtain information about the
secret key.

\emph{Conclusion --- } In this Letter we demonstrated that if one
allows {an} adversary a limited access to {the} random sources
used by   the  {communicating} parties {then} the security of QKD
protocols {is} be completely compromised. This is the case {for}
almost all known {QKD} protocols that use part of the {data set to
test for an adversary.} In such {instances,} the adversary is able
to restrict  the test sample efficiently. The obvious defence
against such an attack is to increase the number of {test states
to a significant} linear portion of the raw key. {This would,
however, {profoundly} decrease the length  of the secret key.}

\emph{Acknowledgements --- }We acknowledge the support of the
Czech Science Foundation GA{\v{C}}R projects P202/12/1142 and
P202/12/G061, as well as projects CE SAS QUTE and VEGA 2/0072/12.
MPl acknowledges the support of SoMoPro project funded under FP7
(People) Grant Agreement no 229603 and by South Moravian Region.
CW acknowledges support from the Marie Curie Actions programme.

\bibliographystyle{ieeetr}
{}

\end{document}